\documentclass{mn2e}
\usepackage{epsfig,natbib2,natbibmnfix}
\usepackage{amssymb}
\usepackage{multirow}
\usepackage{multicol}
\usepackage[varg]{txfonts}
\usepackage{color}
\newcommand{\fref}[1]{Fig.~\ref{#1}}
\newcommand{\tref}[1]{Table~\ref{#1}}
\newcommand{\eref}[1]{equation~\ref{#1}}

\def \tc{t_{\mathrm{cool}}}
\def \cs{c_{\mathrm{s}}}
\def \msolaryr{\; M_{\odot} \; \mathrm{yr}^{-1}}

\begin{document}


\title[Opacity effects in protoplanetary discs]{The effects of opacity on
  gravitational stability in protoplanetary discs}  

\author[P.~Cossins, G.~Lodato \& C.~Clarke] 
{
\parbox{5in}{Peter Cossins$^1$, Giuseppe Lodato$^{1,2}$ and  Cathie Clarke$^3$}
\vspace{0.1in} 
 \\ $^1$ Dept. of Physics \& Astronomy, University of Leicester, Leicester LE1
 7RK UK  
 \\ $^2$ Dipartimento di Fisica, Universit\`{a} Degli Studi di Milano, Via 
 Celoria 16, 20133, Milano, Italia 
 \\ $^3$ Institute of Astronomy, Madingley Road, Cambridge CB3 0HA UK
}

\maketitle

\begin{abstract}
In this paper we consider the effects of opacity regimes on the stability of
self-gravitating protoplanetary discs to fragmentation into bound objects.
Using a self-consistent 1-D viscous disc model, we show that the ratio of
local cooling to dynamical timescales $\Omega \tc$ has a strong dependence on
the local temperature.  We investigate the effects of temperature-dependent
cooling functions on the disc gravitational stability through controlled
numerical experiments using an SPH code.  We find that such cooling functions
raise the susceptibility of discs to fragmentation through the influence of
temperature perturbations -- the average value of $\Omega \tc$ has to
increase to prevent local variability leading to collapse.
We find the effects of temperature-dependence to be most significant in the
`opacity gap' associated with dust sublimation, where the average value of
$\Omega \tc$ at fragmentation is increased by over an order of magnitude.  We
then use this result to predict where protoplanetary discs will fragment into
bound objects, in terms of radius and accretion rate.  We find that without
temperature dependence, for radii $\lesssim 10 AU$ a very large accretion rate
$\sim 10^{-3} \msolaryr$ is required for fragmentation, but that this is
reduced to $10^{-4} \msolaryr$ with temperature-dependent cooling.  We also
find that the stability of discs with accretion rates $\lesssim 10^{-7}
\msolaryr$  at radii $\gtrsim 50 AU$ is enhanced by a lower background
temperature if the disc becomes optically thin.   
\end{abstract}

\begin{keywords}
{accretion, accretion discs -- gravitation -- instabilities -- planets and
  satellites: formation} 
\end{keywords}

\footnotetext[1]{E-mail: peter.cossins@astro.le.ac.uk}

\section{Introduction}
\label{intro}

The formation of planets within protoplanetary discs is a subject that
attracts considerable interest, with two main competing schools of thought.
The core accretion-gas capture model \citep{Lissauer93, LissauerS07, Klahr08}
posits  hierarchical growth, with the collisional coagulation of dust grains
initially leading to centimetre-sized particles, and thence on to
planetesimals and rocky planets.  Once a critical mass is reached, it is then
possible to accrete a gaseous envelope and hence form giant Jupiter-like
planets.  Various observations have successfully confirmed this mode of planet
formation, for example \citet{Marcyetal05}, \citet{Dodson-RobinsonB09}.   

However, this model cannot explain all the available observations.
\citet{KennedyK08} show that beyond approximately 20 AU, the timescales for
giant planet formation via core accretion exceed the expected disc lifetime of
approximately 10 Myr, implying that no planets should be detected in this
region.  However, recent observations of HR8799 with the Keck and Gemini
telescopes have produced direct images of giant planets (5 - 13
$M_{\mathrm{J}}$) orbiting at radii of up to $\sim 70$ AU \citep{Maroisetal08}.
Similar observations of other systems (e.g. $\beta$ Pic b
\citep{Lagrangeetal09} and Formalhuat \citep{Kalasetal08}) and theoretical work
on the formation of 2MASS1207b \citep{LodatoDC05} have suggested that there is
another mechanism for planet formation at work, and this is thought to be the
effect of gravitational instabilities within the protoplanetary discs
themselves.  

In protoplanetary discs where the self-gravity of the gas is dynamically
important, direct gravitational collapse of locally Jeans-unstable
over-densities within the disc \citep{Boss97, Boss98, Durisenprpl} would also
produce giant planets very rapidly, on the local dynamical timescale.  A
similar process of gravitational instability leading to local collapse is a
strong candidate for the formation of stellar discs around Active Galactic
Nuclei (AGN) \citep{NayakshinCS07} and those observed in our own Galactic
Centre \citep{LevinB03, NayakshinC05}, and in the context of protostellar
discs may furthermore be responsible for the formation of brown dwarves and
other low mass stellar companions \citep{StamatellosHW07}.    

The emergence of the gravitational instability within a disc is governed by
the parameter $Q$ \citep{Toomre64}, which for a gaseous Keplerian disc is
given by  
\begin{equation}
  Q = \frac{\cs \Omega}{\pi G \Sigma}.
  \label{Q}
\end{equation}
This encapsulates the balance between the stabilising effects of rotation
($\Omega(R)$ is the angular frequency at radius $R$) and thermal pressure
($\cs (R)$ is the sound speed) and the destabilising effect of the
disc self-gravity via the surface density $\Sigma(R)$.  When $Q \lesssim 1$
the instability is initiated, leading to the presence of spiral density waves
within the disc which, depending on the local cooling rate, may persist in a
self-regulated quasi-stable state \citep{Gammie01, LodatoR04} or may fragment
into bound clumps \citep{JandG03,   RiceLA05}.  Discs that are sufficiently
cool are therefore expected to be susceptible to the gravitational
instability, and it is thought that at least in the early stages of stellar
evolution, many discs enter the self-gravitating phase \citep{Hartmannbook}. 

Once the gravitational instability is initiated, heat is input to the disc on
the dynamical timescale through the passage of spiral compression/shock waves
\citep{Cossinsetal09}.  Various numerical studies using both 2D and 3D models
of self-gravitating discs have produced the result that, in order to induce
fragmentation, the disc must be able to cool on a timescale faster than a few
times the local dynamical time, $t_{\mathrm{dyn}} = \Omega^{-1}$
\citep{Gammie01, RiceLA05}.  This condition is likely to occur only at
relatively large radii ($\sim 100$ AU), on the assumption that stellar or
external irradiation of the disc is negligible \citep{Rafikov09,
  StamatellosW09}. 

These models have generally used a cooling rate prescribed by using a fixed
ratio between the local cooling ($\tc$) and dynamical ($\Omega^{-1}$) times,
such that  
\begin{equation}
  \Omega \tc = \beta
\end{equation}
for some constant $\beta$ throughout the radial extent of the disc.  Various
authors (e.g. \citealt{Gammie01, RiceLA05}) have found that fragmentation occurs
whenever $\Omega \tc \approx 3 - 7$.  By using a more realistic cooling
framework based on the optical depth, \citet{JandG03} found that the
fragmentation boundary (defined hereafter as the ratio of the cooling to
dynamical timescales, $\Omega \tc$ at fragmentation) may in fact be over an
order of magnitude greater than this, leading to an enhanced tendency towards
fragmentation.  This variation in $\Omega \tc$ they ascribed to the implicit
dependence of the cooling function on the disc opacity, and hence on
temperature.  

Using the opacity tables of \citet{BellLin94} it is clear that the opacity is
a strong function of temperature in certain regimes, and by modelling
protoplanetary discs as optically thick in the Rosseland mean sense, $\Omega
\tc$ shows power law dependencies on both the local temperature and
density. In cases where this dependence is strong, it is therefore possible
that small temperature fluctuations may push the local value of $\Omega \tc$
below the fragmentation boundary, even when the \textit{average} value is
significantly above it.    

In this paper we therefore seek to investigate and clarify the exact
relationship between the fragmentation boundary and the temperature dependence
of $\Omega \tc$, using a Smoothed Particle Hydrodynamics (\textsc{sph}) code
to conduct global, 3D numerical simulations of discs where the cooling time
follows a power-law dependence on the local temperature.  In addition,
various studies have shown that in a quasi-steady state the gravitational
instability may be modelled pseudo-viscously (\citealt{LodatoR05,
  Cossinsetal09, Clarke09, Rafikov09}).  We therefore use the
$\alpha$-prescription of \citet{ShakSun73} and the assumption of local thermal
equilibrium, where    
\begin{equation}
  \Omega \tc = \frac{4}{9} \frac{1}{\gamma (\gamma - 1) \alpha}
  \label{thermalbalance}
\end{equation}
and $\gamma$ is the ratio of specific heats, to construct an analytical model
of the opacity regimes present within a marginally gravitationally stable
disc.  From this we can therefore predict analytically if and where such discs
would become prone to fragmentation, and also compare these results to those
from the more complex simulations where radiative transfer is modelled, such
as \citet{Boley09} and \citet{StamatellosW09}. 

The structure of this paper is therefore as follows.  In Section 2 we
discuss some of the theoretical results relevant to protoplanetary discs, and
introduce a simplified cooling function derived from the various opacity
regimes.  We further consider the effects we expect these cooling
prescriptions to have on the susceptibility of protoplanetary discs to
fragmentation.  In Section 3 we briefly outline the numerical modelling
techniques used in our simulations and detail our initial conditions.  In
Section 4 we present the results from these simulations, before proceeding to
collate these with the analytical predictions in Section 5.  Finally in
Section 6 we discuss the ramifications of our work and the conclusions that may
be drawn from it. 

\section{Theoretical Results}
\label{theory}

In this section we derive analytical results for the dependence of the
cooling timescale $\tc$ on temperature and density, such as we might expect to
find in a quasi-gravitationally stable protoplanetary disc environment.  We
also consider analytically the effects that a (specifically) temperature
dependent cooling time will have on the stability of such a disc to
fragmentation. 

\subsection{$\Omega \tc$ in the Optically Thick Regime}
\label{opacitytheory}
As in the case of \citet{Gammie01}, we may start from the following basic
equations:
\begin{eqnarray}
  \tc    & =       & \frac{U \Sigma}{\Lambda},  
  \label{tcool}\\
  \tau   & \approx & \rho H \kappa,  
  \label{tau} \\
  \Sigma & =       & 2 \rho H,
  \label{Sigma} \\
  \cs^2  & =       & \frac{\gamma \mathcal{R} T}{\mu},
  \label{cs2}
\end{eqnarray}
where $U$ is the specific internal energy, $\Lambda$ is the cooling rate per
unit area, $\tau$ is the optical depth, $\rho$ is the (volume) density, $H =
\cs / \Omega$ is the disc scale height, $\kappa$ is the opacity, $\gamma$ is
the ratio of specific heats, $\mathcal{R} = k/m_{\mathrm{H}}$ is the universal
gas constant ($k$ being the Boltzmann constant and $m_{\mathrm{H}}$ the mass
of a hydrogen atom), $T$ is the local mid-plane temperature and $\mu$ is the
mean molecular weight of the gas.  Note that the factor of two in \eref{Sigma}
arises from there being two faces of the disc from which to radiate.   

In the case where the disc is optically thick (in terms of the Rosseland
mean), then the cooling rate per surface area $\Lambda$ may be given as 
\begin{equation}
  \Lambda = \frac{16 \sigma T^4}{3 \tau},
\end{equation}
where $\sigma$ is the Stefan-Boltzmann constant.  We note that this is
strictly valid only in the case where energy is transported radiatively
within the disc --- convective transport or stratification within the disc will
alter this relationship (see for example \citealt{Rafikov07}).  For the purely
radiative case, the vertical temperature structure of the disc is therefore
accounted for via this formalism, and is characterised by the midplane
temperature $T$ and the optical depth $\tau$.  In order to prevent divergence
of this cooling function at low optical depths and to interpolate smoothly
into the optically thin regime, others including \citet{JandG03} and
\citet{RiceA09} have used a cooling function of the form   
\begin{equation}
  \Lambda = \frac{16 \sigma T^4}{3} \left(\tau + \frac{1}{\tau} \right)^{-1},
  \label{Lambda}
\end{equation}
which becomes directly proportional to the optical depth in the optically thin
limit.  In general however, we find that discs only become optically thin at
large radii, and that this correction is therefore only relevant to the case
where the cooling is dominated by ices.

Furthermore, we note that for systems where the stellar mass dominates over
that of the disc, the density $\rho$ may be approximated by   
\begin{equation}
  \rho \approx \frac{M_{*}}{2 \pi R^3 Q}
  \label{rho}
\end{equation}
where $M_{*}$ is the mass of the central star and $R$ the radial distance from
the central star, and therefore we have $\Omega^2 = 2 \pi G \rho Q$ in the
case of Keplerian rotation, with $G$ being the universal gravitation
constant. Recalling also that $\cs^2 = U \gamma (\gamma - 1)$,
equations \ref{tcool} -- \ref{rho} may be rearranged to show that in the
optically thick case, the ratio of cooling to dynamical times should be  
\begin{equation}
  \Omega \tc = \frac{3 \mathcal{R}^2}{8 \sigma \sqrt{2 \pi G}}
  \frac{\gamma}{\gamma - 1} \frac{\kappa}{\mu^2} \; Q^{-1/2} \; \rho^{3/2} \;
  T^{-2}. 
  \label{thickOtc}
\end{equation}

\citet{BellLin94} found that the opacity can be reasonably well approximated
by power-law dependencies on temperature and density, such that 
\begin{equation}
  \kappa = \kappa_{0} \; \rho^{a} \; T^{b}.
  \label{opacity}
\end{equation}
Specific values of $a$, $b$ and $\kappa_{0}$ apply for each opacity regime,
such that the value of $\kappa$ varies continuously over the regime
boundaries.  Using these approximations, we find that the $\Omega \tc$ value
for the various opacity regimes can be given by
\begin{equation}
  \Omega \tc = \frac{3 \mathcal{R}^2}{8 \sigma \sqrt{2 \pi G}}
  \frac{\gamma \kappa_{0}}{\mu^2(\gamma - 1)} \; Q^{-1/2} \; \rho^{a + 3/2} \;
  T^{b-2}. 
  \label{Omtc}
\end{equation}
For each opacity regime, the constant $\kappa_{0}$, the exponents $a$ and $b$,
the transition temperatures between the regimes and the functional dependence
of $\Omega \tc$ on temperature and density are given in \tref{opacities}.
It should be noted that for the purposes of these tables the temperature and
density should be measured in cgs units.  

\begin{table*}
  \caption{Details of the various optical regimes by type, showing the
    transition temperatures and the functional dependence of $\Omega \tc$ on
    the temperature and density in the optically thick regime.  Note that all
    values are quoted in cgs units.  See \citet{BellLin94} for further
    details.}
  \begin{center}
    \begin{tabular}{ccccccc}
      \hline
      \multirow{2}{*}{\textbf{Opacity Regime}} &
      \multirow{2}{*}{$\mathbf{\kappa_{0}}$ (cm$^2$ g$^{-1}$)} &
      \multirow{2}{*}{$\mathbf{a}$} & 
      \multirow{2}{*}{$\mathbf{b}$} &
      \multicolumn{2}{c}{\textbf{Temperature Range } ($\mathbf{K}$)} & 
      \multirow{2}{*}{\textbf{Dependence of $\Omega \tc$}} \\
      & & & & \textbf{From} & \textbf{To} & \\
      \hline
      Ices                       & $2 \times 10^{-4}$  & 0   & 2   & 0
      & $166.810$ &  $\rho^{3/2}$        \\  
      Sublimation of Ices        & $2 \times 10^{16}$  & 0   & -7  & $166.810$
      & $202.677$ & $\rho^{3/2} \; T^{-9}$ \\ 
      Dust Grains                & $1 \times 10^{-1}$  & 0   & 1/2 & $202.677$
      & $2286.77 \; \rho^{2/49}$ & $\rho^{3/2} \; T^{-5/2}$ \\ 
      Sublimation of Dust Grains & $2 \times 10^{81}$  & 1   & -24 & $2286.77
      \; \rho^{2/49}$ & $2029.76 \; \rho^{1/81}$ & $\rho^{5/2} \; T ^{-26}$ \\
      Molecules                  & $1 \times 10^{-8}$  & 2/3 & 3   & $2029.76
      \; \rho^{1/81}$ & $10000.0 \; \rho^{1/21}$ & $\rho^{13/6} \; T^{1}$ \\
      Hydrogen scattering        & $1 \times 10^{-36}$ & 1/3 & 10  & $10000.0
      \; \rho^{1/21}$ & $31195.2 \; \rho^{4/75}$ & $\rho^{11/6} \; T^{8}$ \\
      Bound-Free \& Free-Free    & $1.5 \times 10^{20}$& 1   &-5/2 & $31195.2
      \; \rho^{4/75}$ & $1.79393 \times 10^{8} \; \rho^{2/5}$ & $\rho^{5/2} \;
      T^{-9/2}$ \\ 
      Electron scattering        & 0.348              & 0 & 0 & $1.79393
      \times 10^{8} \; \rho^{2/5}$ &  -------- & $\rho^{3/2} \; T^{-2}$ \\ 
      \hline
    \end{tabular}
  \end{center}
  \label{opacities}
\end{table*}

\subsection{Effects of Temperature Dependence on Fragmentation}
We now specifically consider the effects of temperature fluctuations on the
stability of a disc to fragmentation, using a simplified cooling prescription
derived from a consideration of \eref{Omtc}.    

In the previous section it was noted that the ratio of the local cooling and
dynamical times $\Omega \tc$ has a direct dependence on the local mid-plane
temperature $T$. Given that (from \tref{opacities}) this dependence is generally
much stronger than that on density, it is physically reasonable to consider a
simplified cooling function where we only include the effects of temperature,
and where we \textit{define} the cooling time via the relationship  
\begin{equation}
  \Omega \tc = \beta \left( \frac{T}{\bar{T}} \right)^{-n},
  \label{modelcooling}
\end{equation}
for some general value of the cooling exponent $n$ and cooling parameter
$\beta$.  Here $\bar{T}$ is the azimuthally averaged mid-plane temperature $T$
in thermal equilibrium, and thus we see that when thermal equilibrium is
reached, the average cooling timescale is expected to reduce to $\left< \Omega
\tc \right> \approx \beta$, with a fragmentation boundary $\beta_{n}$
associated with each value of $n$.  In particular, with $n = 0$, at
fragmentation we have $\Omega \tc = \left< \Omega \tc \right> = \beta_{0}$,
which \citet{Gammie01}, \citet{RiceLA05} and others have found to be in the
range $3 - 7$.    

In the case of temperature dependent cooling (where $n \neq 0$), if the
equilibrium value of the cooling parameter $\beta > \beta_{0}$, the disc may
still fragment due to temperature fluctuations leading to a short term (relative
to the dynamical timescale) decrease in the instantaneous value of $\beta$ to
less than the threshold value.  For a power-law index $n$, in order to
calculate the value $\beta_{n}$ of the equilibrium cooling parameter below
which fragmentation occurs, we make the assumption that fragmentation takes
place wherever the instantaneous value of $\Omega \tc$ is held at or below the
critical value $\beta_{0}$ for longer than a dynamical time,
\textit{independent} of the mechanism by which the cooling is effected.  If we
therefore consider temperature fluctuations such that $T = \bar{T} + \delta
T$, we find that at the fragmentation boundary   
\begin{equation}
  \beta_{0} = \beta_{n} \left( 1 + \frac{\delta T}{\bar{T}} \right)^{-n}.
  \label{boundary}
\end{equation}

In \citet{Cossinsetal09} for the case where $M_{\mathrm{disc}} / M_{*} = 0.1$
we found that on average the strength of the surface density perturbations
$\delta \Sigma / \bar{\Sigma}$ can be linked to the strength of the cooling
through the following relationship, 
\begin{equation}
  \left< \frac{\delta \Sigma}{\bar{\Sigma}} \right> \approx \frac{1}{\left<
    \Omega \tc \right> ^{1/2}}, 
  \label{Sigmafluc}
\end{equation}
where angle brackets denote the RMS value.  In a similar manner we may say
that 
\begin{equation}
  \left< \frac{\delta T}{\bar{T}} \right> = \frac{k}{\left< \Omega \tc
    \right>^{1/2}}, 
  \label{Tfluc}
\end{equation}
where $k$ is to be defined empirically.   At fragmentation therefore we have 
\begin{equation}
  \left< \frac{\delta T}{\bar{T}} \right> = \frac{k}{\beta_{n}^{1/2}},
\end{equation}
noting that by construction for a given index $n$, at fragmentation $\left<
\Omega \tc \right> = \beta_{n}$.  Combining this with \eref{boundary} we find
that in the case where the cooling is allowed to vary with temperature as per
\eref{modelcooling}, the fragmentation boundary $\beta_{n}$ satisfies
the following equation;  
\begin{equation}
  \beta_{0} = \beta_{n} \left( 1 + \frac{k}{\beta_{n}^{1/2}} \right)^{-n}.
  \label{betaboundary}
\end{equation}
This implicit equation can therefore be solved to find the value of the
fragmentation boundary $\beta_{n}$ for all $n \gtrsim -2$ (below this
$\beta_{n}$ becomes undefined), as shown later in \tref{opacityfrag}.


\section{Numerical Set Up}
\label{setup}

\subsection{The SPH code}
All of the simulations presented hereafter were performed using a 3D smoothed
particle hydrodynamics (\textsc{sph}) code, a Lagrangian hydrodynamics code
capable of modelling self-gravity (see for example, \citealt{Benz90},
\citealt{Monaghan92}).  The code self-consistently incorporates the so-called
$\nabla h$ terms to ensure energy conservation, as described in
\citet{SpringelH02}, \citet{PriceM07}.  All particles evolve according to
individual time-steps governed by the Courant condition, a force condition
\citep{Monaghan92}, an integrator limit \citep{BateBP95} and an additional
condition that ensures the local timestep is always less than the local
cooling time.  

We have modelled our systems as a single point mass (onto which gas
particles may accrete if they enter within a  given sink radius and satisfy
certain boundness conditions --- see \citealt{BateBP95}) orbited by 500,000
\textsc{sph} gas particles; a set up common to many other \textsc{sph}
simulations of such systems (e.g. \citet{Riceetal03, LodatoR04,  LodatoR05,
  Harperclark07, Cossinsetal09}) but at a higher resolution than most.
The central object is free to move under the gravitational influence of the
disc.    

In common with many other simulations where cooling is being investigated
(\citet{Gammie01, LodatoR05, Cossinsetal09} for example) we use a simple
implementation of the following form;
\begin{equation} 
  \frac{du_{{i}}}{dt} = - \frac{u_{i}}{t_{\mathrm{cool},i}},
\end{equation}
where $u_{i}$ and $t_{\mathrm{cool},i}$ are the specific internal energy and
cooling time associated with each particle respectively.  The cooling time is
allowed to vary with the particle temperature $T_{i}$ in such a manner that 
\begin{equation}
  \Omega_{i} t_{\mathrm{cool},i} = \hat{\beta} \left(
  \frac{T_{i}}{\bar{T}} 
  \right)^{-n},
  \label{cooling}
\end{equation}
where $\Omega_{i}$ is the angular velocity of the particle, $\bar{T}$ is
the equilibrium temperature, and $\hat{\beta}$ and $n$ are input values held 
constant throughout any given simulation.  Given that $T \sim \cs^{2}$,
\eref{Q} shows that for a given value of the surface density $\Sigma$ this is
equivalent to 
\begin{equation}
  \Omega_{i} t_{\mathrm{cool},i} = \hat{\beta} \left(
  \frac{Q_{i}}{\bar{Q}} 
  \right)^{-2n},
  \label{simcooling}
\end{equation}
where again $Q_{i}$ is the value of the $Q$ parameter evaluated at each
particle, and $\bar{Q}$ is the expected equilibrium value of $Q$, which we take
to be 1 throughout.  Note that \textit{a priori} we do not know exactly what
the equilibrium value of $Q$ will be once the gravitational instability has
saturated.  Indeed as we shall see this turns out to be slightly greater than
unity, but still such that $Q \approx 1$.  The \textit{effective} value of the
cooling parameter is given by  
\begin{equation}
  \beta = \hat{\beta} Q^{-2n},
  \label{beta_def}
\end{equation}
where $Q$ is the actual value to which the simulations settle.  Since we are
exploring relatively large values of $n$, $\beta$ can vary significantly from
our input value $\hat{\beta}$ for even small changes in $Q$. 

Finally we calculate the equivalent surface density
$\Sigma_{i}$ (and thus $Q_{i}$) at the radial location of each particle
$R_{i}$ by dividing up the disc into (cylindrical) annuli, calculating the
surface density for each annulus, and then interpolating radially to obtain
$\Sigma_{i}(R_{i})$.  To prevent boundary effects, for simulations where $n >
1.0$ the temperature dependent effects are limited to an annulus $15 \leq R \leq
20$ (in code units --- note that initially $R_{\mathrm{in}} = 0.25$ and
$R_{\mathrm{out}} = 25.0$).  At other radii we keep $\Omega \tc = 8$, a value
chosen to suppress fragmentation in regions outside the annulus of interest
(see for instance \citealt{Alexanderetal08}). 

All simulations have been run with the particles modelled as a perfect gas, with
the ratio of specific heats $\gamma = 5/3$.  Heat addition is allowed for
via $P\mathrm{d}V$ work and shock heating.  Artificial viscosity has been
included through the standard \textsc{sph} formalism, with
$\alpha_{\mathrm{SPH}} = 0.1$ and $\beta_{\mathrm{SPH}} = 0.2$ ---  although
these values are smaller than those commonly used in \textsc{sph} simulations,
this limits the transport and heating induced by artificial viscosity.  As
shown in \citet{LodatoR04}, with this choice of parameters the transport of
energy and angular momentum due to artificial viscosity is a factor of 10
smaller than that due to gravitational perturbations, while we are still able
to resolve the weak shocks occurring in our simulations.   
  
By using the cooling prescription outlined above in \eref{simcooling}, the
rate at which the disc cools is governed by the dimensionless parameters $Q$,
$\hat{\beta}$ and $n$, and the cooling is thereby implemented scale free.  The
governing equations of the entire simulation can therefore likewise be recast
in dimensionless form.  In common with the previous \textsc{sph} simulations
mentioned above, we define the unit mass to be that of the central object --
the total disc mass and individual particle masses are therefore expressed as
fractions of the central object mass.  We can self-consistently define an
arbitrary unit (cylindrical) radius $R_{0}$, and thus, with $G = 1$, the unit
timestep is the dynamical time $t_{\mathrm{dyn}} = \Omega^{-1}$ at radius $R =
1$.    

\subsection{Initial conditions}
All our simulations model a central object of mass $M_{*}$, surrounded
by a gaseous disc of mass $M_{\mathrm{disc}} = 0.1 M_{*}$.  We have used an
initial surface density profile $\Sigma \sim R^{-3/2}$, which implies that in
the marginally stable state where $Q \approx 1$, the disc temperature profile
should be approximately flat for a Keplerian rotation curve.  Since the
surface density evolves on the viscous time $t_{\mathrm{visc}} \gg
t_{\mathrm{dyn}} = \Omega^{-1}$ this profile remains roughly unchanged
throughout the simulations.   Radially the disc extends from $R_{\mathrm{in}}
= 0.25$ to $R_{\mathrm{out}} = 25.0$, as measured in the code units described
above.  The disc is initially in approximate hydrostatic equilibrium in a
Gaussian distribution of particles with scale height $H$. The azimuthal
velocities take into account both a pressure correction \citep{LodatoRNC07}
and the enclosed disc mass.  In both cases, any variation from dynamical
equilibrium is washed out on the dynamical timescale.  

The initial temperature profile is $\cs^{2} \sim R^{-1/2}$ and is
such that the minimum value of the Toomre parameter $Q_{\mathrm{min}} = 2$
occurs at the outer edge of the disc. In this manner the disc is initially
gravitationally stable throughout.  Note that the disc is \textit{not}
initially in thermal equilibrium -- heat is not input to the disc until
gravitational instabilities are initiated.  

\subsection{Simulations run}
Since our simulations use a slightly different surface density profile to that
used by previous authors ($\Sigma \sim R^{-3/2}$, cf. $\Sigma \sim
R^{-1}$ in \citealt{RiceLA05}, $\Sigma \sim R^{-7/4}$ in \citealt{Riceetal03})
we initially ran five simulations at various values of $\beta$ with the
cooling exponent $n$ set equal to zero, to find the fragmentation boundary in
the case where the cooling is independent of temperature.  Thereafter,
simulations were run at various $\beta$ values as $n$ was incremented up to $n
= 3$, to ascertain the fragmentation boundary in each case.  A summary of all
the simulations run is given in Table \ref{simulations}. 
\begin{table}
  \caption{Table of simulations run for various values of the cooling exponent
    $n$ and rate $\beta$.  Note that since many of these simulations were run
    concurrently, there is a degree of overlap in the $\beta$ values used.}
  \begin{center}
    \begin{tabular}{|c|c|}
      \hline 
      \textbf{Exponent} ($n$) &\textbf{Input cooling parameter}
      ($\hat{\beta}$) \\ 
      \hline
      0.0 & 3, 4, 4.5, 5, 6\\
      0.5 & 4, 4.5, 5, 5.5, 6\\
      1.0 & 3, 4, 5, 6, 7, 8, 9, 10\\
      1.5 & 7, 8, 9, 10, 11\\
      2.0 & 10, 11, 12, 13, 14, 15, 16, 17, 18\\
      3.0 & 20, 22.5, 25, 27.5, 30, 32.5, 35, 37.5, 40\\
      \hline
      \end{tabular}
  \end{center}
  \label{simulations}
\end{table}


\section{Simulation Results}
\label{results}

\subsection{Detecting Fragmentation}
\begin{figure}
  \centering
  \includegraphics[width=20pc]{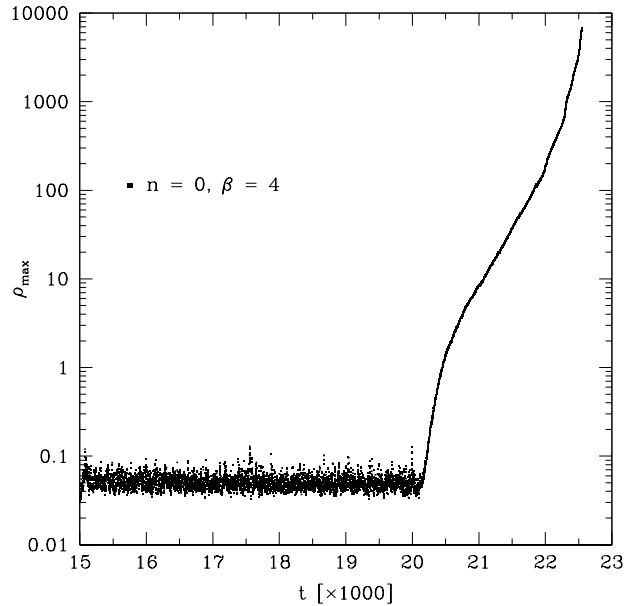}
  \caption{Maximum density plot showing the characteristic rise due to fragment
  formation, seen here for the simulation where $\beta = 4.0$, $ n=0$ (where
  the cooling is independent of temperature).  There is clear evidence of
  fragment formation at $t \approx 20,000$, with both density and time being
  shown in code units.} 
  \label{fragmentation}
\end{figure}

First of all it is useful to explain how fragmentation has been detected in
our simulations. Throughout all the numerical simulations run, the maximum
density over all particles has been tracked as a function of elapsed time.  In
the case of a non-fragmenting disc, the maximum always occurs at the inner
edge of the disc (as would be expected), and is relatively stable over time.
However, once a fragment forms, this maximum density (now corresponding to the
radius at which the fragment forms) rises exponentially, on its own dynamical
timescale.  An example is shown in \fref{fragmentation}, and the various
changes in gradient correspond to various fragments at different radii (and
thus with differing growth rates) achieving peak density.  A similar increase
in the central density of proto-fragments is observed in
\citet{StamatellosW09a}, although the timescales differ due to the use of
different equations of state.

This rise in the maximum density has therefore been used throughout as a
tracer of fragment formation, and the evolution has been followed until the
fragments are at least four orders of magnitude greater than the original
peak density.

\subsection{Averaging Techniques}
\label{averaging}
\begin{figure}
  \centering
  \includegraphics[width=20pc]{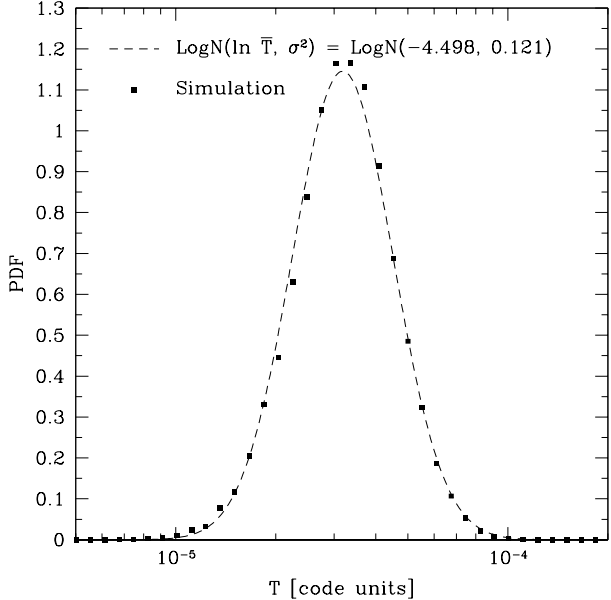}
  \caption{Distribution of particle temperatures for $16.25 \leq R \leq 18.75$,
    and a predicted log-normal distribution based on the same data.  The two
    are equal to within approximately 1\%.}
  \label{logNormalT}
\end{figure}

Throughout the following analysis, we have defined the average value of a
(strictly positive) quantity, which we denote by an overbar, as the
\textit{geometric} mean of the particle quantities.  The reason for this is we
find that in the ``gravo-turbulent'' equilibrium state properties such as the
temperature, density and Q value are log-normally distributed. This is shown
for example in \fref{logNormalT}, where the temperature data from the
simulation match a predicted log-normal distribution to within one percent.
(Note the reduced radial range to reduce the effect of the inherent gradual
reduction in temperature with radius.)  The geometric mean being precisely
equivalent to the exponential of the arithmetic mean of the logged values,
this process recovers the mean value of the normal distribution of $\ln T$.  

Similarly, to calculate the perturbation strengths (e.g $\delta A / \bar{A}$
for some quantity $A$) we note that 
\begin{equation}
  \frac{\delta A}{\bar{A}} \approx \frac{\mathrm{d}A}{A} = \mathrm{d} \ln A.
\end{equation}
The RMS value of $\delta A / \bar{A}$ is then equivalent to the standard
deviation of $\ln A$, which again can be recovered directly from the
log-normal distribution.  Referring again to \fref{logNormalT} we therefore see
that $\bar{T} = 10^{-4.498} = 3.177 \times 10^{-5}$ (in code units), and
that $\delta T / \bar{T} = \sigma = 0.348$.
  
\subsection{Equilibrium States}
First of all, we need to determine the exact value of the fragmentation
boundary in the case where $n = 0$ (and thus where $\beta = \hat{\beta}$),
which we denote by $\beta_{0}$.  As seen in \tref{simulations}, simulations
were run at various values $3.0 \leq \beta \leq 6.0$, and we find that the
boundary lies between 4.0 and 4.5.  We therefore take the critical value as
being the midpoint, such that $\beta_{0} = 4.25$. 

\begin{figure}
  \centering
  \includegraphics[width=20pc]{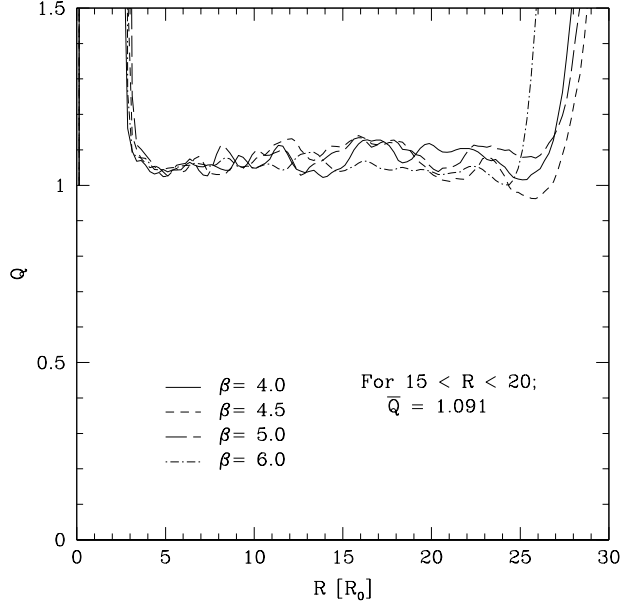}
  \caption{Plot of $Q$ against radius for various values of $\beta$ in the
    temperature independent case $n = 0$.  For the fragmenting cases ($\beta <
    4.25$) the values shown are from immediately prior to fragmentation.} 
  \label{Qn0}
\end{figure}

Continuing with the $n = 0$ case, we find throughout that the value of $Q$ to
which the simulations settle is slightly above unity.  The steady state values
(time averaged over 1000 timesteps) are shown for various $\beta$ values in
\fref{Qn0}, and we see that the average $Q$ value is approximately 1.091,
where we have averaged over both $\beta$ and radius (where $15 \leq R \leq
20$, for comparison with simulations with higher $n$).  We further note that
there is scatter of $\sim 10\%$ about this average, and (although not shown)
this is equally true of the simulations where $n > 0$.

Note then that for large $n$ the effective value of the cooling parameter
$\beta$ at any given radius may be substantially different from the numerical
input value $\hat{\beta} = \beta Q^{2n}$ (see \eref{beta_def}) that we use to
characterise the cooling law.  In order to determine the fragmentation
boundary with any accuracy, we therefore need to consider the true value of
$\beta$ rather than the input value $\hat{\beta}$.

\subsection{Cooling Strength and Temperature Fluctuations}
\begin{figure}
  \centering
  \includegraphics[width=20pc]{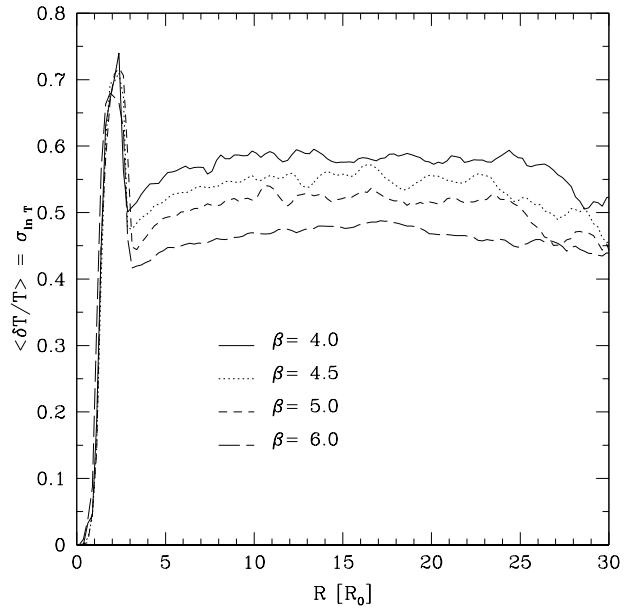}
  \caption{Plot showing the strength of temperature perturbations within the
    disc as a function of radius and $\beta$ for the temperature independent
    case, where $n = 0$.} 
  \label{dtotn0}
\end{figure}

In order to characterise the fragmentation boundary, it is necessary that we
validate the assumption encompassed by \eref{Tfluc}, that the temperature
perturbation strength is correlated to that of the applied cooling.  Using
the method outlined above in section \ref{averaging}, for each simulation we
can calculate azimuthally averaged RMS values for the strength of the
temperature fluctuations, which we denote by $\left< \delta T /\bar{T} \right>
$.  Where $n = 0$, these temperature perturbations are plotted as a function
of radius for various values of $\beta$ in \fref{dtotn0}, where we see that
there is a systematic decrease in the perturbation strength with increasing
$\beta$, and also that the perturbation strength is almost constant with
radius across the self-regulating region ($5 \lesssim R \lesssim 25$) of the
disc.  Using \eref{Tfluc} we can therefore calculate an empirical value for
$k$, and hence averaging both radially (for $15 \leq R \leq 20$ as before) and
over the available values of $\beta$ we find $k = 1.170$ where $n = 0$.  
 
Furthermore, we note that in the temperature dependent case (where $n \neq
0$), by construction the average value $\left< \Omega \tc \right>$ is simply the
effective value of the cooling strength, $\beta$.  We can therefore calculate
the value of $k$ for cases where $n \neq 0$, and we find that again $k$ remains
constant both with the index $n$ and with radius.  Hence we take the value
of $k$ to be 1.170, as in the $n = 0$ case, and empirically we may therefore say
that on average  
\begin{equation}
  \left< \frac{\delta T}{\bar{T}} \right> = \frac{1.170}{\sqrt{\beta}},
  \label{kempirical}
\end{equation}
for all $n$.

\subsection{The Fragmentation Boundary}
\begin{figure}
  \centering
  \includegraphics[width=20pc]{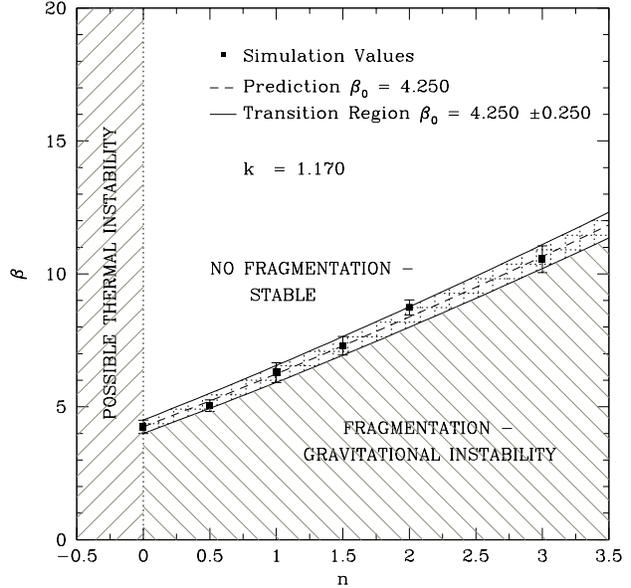}
  \caption{Plot of $\beta_{n}$ at fragmentation for various values of $n$.
    The error bars correspond to the greatest non-fragmenting and smallest
    fragmenting values of $\beta$ found in the simulations, and the
    cross-hatched transition region represents uncertainty in the exact value
    of $\beta_{0}$.  Note also that where $n < 0$ discs may become thermally
    unstable.}
  \label{beta_n}
\end{figure}

We are now in a position to predict empirically the fragmentation boundary in
the case where $n \neq 0$, and to compare this directly with the results of
our simulations.  \tref{results_table} shows the fragmentation boundary
$\beta_{n}$ as obtained from our simulations, where once again it is taken as
the average of the highest fragmenting and lowest non-fragmenting values of
$\beta$ simulated.  We find that as expected, there is indeed a rise in the
fragmentation boundary as the dependence of the cooling on temperature
increases.  This variation of the fragmentation boundary is shown against the
cooling exponent $n$ in \fref{beta_n}, (where the error bars show the upper and
lower bounds from \tref{results_table}) along with predicted values generated
using the following empirically defined implicit relationship 
\begin{equation}
  \beta_{0} = \beta_{n} \left( 1 + \frac{1.170}{\sqrt{\beta_{n}}} \right)^{-n},
  \label{betan_empirical}
\end{equation}
where we have used $\beta_{0} = 4.25$.  Clear from this plot is the fact that
the predictions are a very good match to the data observed, and our
theoretical model, in which the increased tendency for fragmentation is due to
the effects of temperature fluctuations on the cooling rate, is therefore
valid.  The transition zone shown is bounded by curves corresponding to
predictions using $\beta_{0} = 4.00$ and $4.50$, the upper and lower bounds
for $\beta_{0}$ we obtained through our simulations.  

\begin{table}
  \caption{Table showing the fragmentation boundaries obtained from the
    simulations.  The central columns show respectively the highest
    fragmenting and lowest non-fragmenting values of $\beta$ simulated, with
    $\beta_{n}$ being the midpoint of these.  Throughout, $\beta$ is calculated
    using \eref{beta_def}.} 
  \begin{center}
    \begin{tabular}{|c|c|c|c|}
      \hline 
      \multirow{2}{*}{\textbf{Exponent} ($n$)} &
      \multicolumn{2}{c}{\textbf{Effective cooling rate} ($\beta$)} &
      \multirow{2}{*}{$\beta_{n}$}\\  
      & Fragmenting & Non-Fragmenting & \\ 
      \hline
       0.0  &  4.000  &  4.500  &  4.250  \\
       0.5  &  4.825  &  5.263  &  5.044  \\
       1.0  &  5.915  &  6.654  &  6.284  \\
       1.5  &  6.949  &  7.644  &  7.296  \\
       2.0  &  8.458  &  9.022  &  8.740  \\
       3.0  & 10.051  &  11.056 & 10.554  \\
      \hline
     \end{tabular}
  \end{center}
  \label{results_table}
\end{table}

\subsection{Statistical Analysis}
\label{statistics}

\begin{figure}
  \centering
  \includegraphics[width=20pc]{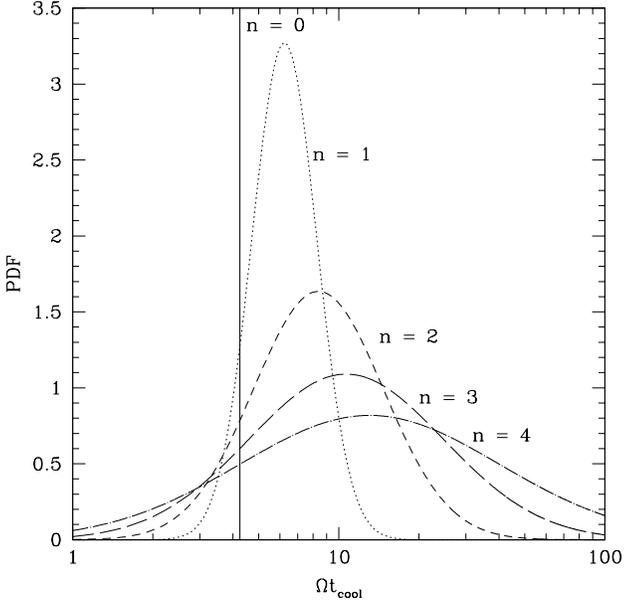}
  \caption{Variation in the distribution of $\ln \Omega \tc$ as a function of
    $n$, clearly showing the increasing width of the distribution with
    increasing $n$.  Note that in the case where $n = 0$ the distribution
    reduces to a $\delta$-function.}
  \label{normaldist}
\end{figure}

The effects of temperature perturbations on the fragmentation boundary can be
neatly illustrated statistically, if we assume that the distribution
of temperatures about the geometric mean  $\ln \bar{T}$ is log-Normal (as found
in our simulations).  Using standard notation we can therefore say that  
\begin{equation}
  \ln T \sim N(\ln \bar{T},\sigma^{2}),
  \label{lognormalT}
\end{equation}
with standard deviation $\sigma$.  By taking logs of \eref{modelcooling} we
further see that   
\begin{equation}
  \ln \Omega \tc = \ln \beta - n \ln T + n \ln \bar{T}.
  \label{logcooling}
\end{equation}
A standard property of the Normal distribution is that for a Normally
distributed random variable $X \sim N(\mu,\sigma^{2})$, the distribution of
$aX + b$ is given by $N(a\mu + b, a^{2} \sigma^{2})$.  Hence from
\eref{logcooling} we see that the distribution of $\ln \Omega \tc$ at
fragmentation is such that 
\begin{equation}
  \ln \Omega \tc \sim N(\ln \beta_{n}, n^{2}\sigma^{2}),
  \label{logOmtc}
\end{equation}
i.e., the distribution of $\ln \Omega \tc$ is centred around $\ln \beta_{n}$
for all $n$, reducing to a $\delta$-function in the limit where $n$ becomes zero
and becoming more spread out as $n$ becomes large.  Thus in order to
counteract the increased width of the distribution, and thus the increased
fraction of the gas that is below the fragmentation threshold, the average
must rise. This is clearly illustrated in \fref{normaldist}, for values of $n$
between 0 and 4, and where $\beta_{n}$ is given in each case by
\eref{betan_empirical} with $\beta_{0} = 4.25$.

\section{Opacity-Based Analytic Disc Models}
\label{analytics}

Having quantified the effects of a temperature-dependent cooling law on the
fragmentation boundary of protoplanetary discs, we are now in a position to
use the known cooling laws for each opacity regime (as given by \eref{Omtc})
to determine the dominant cooling mechanisms throughout the radial range.  We
can therefore also use this to re-evaluate the regions of such discs that are
unstable to fragmentation, in a similar manner to the analysis undertaken by
\citet{Clarke09}. 

In order to do this in a physically realistic manner we must also take into
account the effects of the magneto-rotational instability (MRI), which
operates when the disc becomes sufficiently ionised.  Considering only thermal
ionisation, we assume that the MRI becomes active when the disc temperature
rises above $1000K$ \citep{Clarke09}.  Although estimates of the viscosity
provided through this instability vary (see \citet{Kingetal07} for a summary),
numerical simulations suggest it should be in the range $0.001 \lesssim
\alpha_{\mathrm{MRI}} \lesssim 0.01$ \citep{Wintersetal03, Sanoetal04}.  We
therefore assume that the MRI is the dominant instability in the disc wherever
$T > 1000K$ and the $\alpha$ delivered by the gravitational instability falls
below 0.01.  

To obtain the disc temperature, we note that equations \ref{thermalbalance},
\ref{cs2}, \ref{rho}, \ref{thickOtc} and \ref{opacity} self-consistently allow
the disc properties to be evaluated for any given stellar mass $M_{*}$,
mass accretion rate $\dot{M}$ and radius $R$, when combined with the relation 
\begin{equation}
  \dot{M} = \frac{3 \alpha \cs^{3}}{GQ}
  \label{Mdot}
\end{equation}
(see for instance \citealt{Clarke09, Rafikov09, RiceA09}).  We can thus derive
the dependence of the disc temperature $T$ on $Q$, $M_{*}$, $R$ and $\dot{M}$,
such that
\begin{equation}
  T = \left[ \frac{32 \sigma}{9 \kappa_{0}} \left( \frac{2 \pi \mu}{G \gamma
      \mathcal{R}} \right)^{\frac{1}{2}} \left( \frac{M_{*}}{2 \pi}
    \right)^{-(a +\frac{3}{2})} Q^{a+1} R^{3a + \frac{9}{2}} \dot{M}^{-1}
    \right]^{\frac{2}{2b - 7}}.  
  \label{T-MdotR}
\end{equation}
Finally, in order to prevent the temperature becoming too low, we assume a
fiducial background temperature for the interstellar medium (ISM) of $10K$
\citep{D'Alessioetal98, Hartmannetal98}.  In this case, we no longer assume
that \eref{thermalbalance} holds, as there is additional heating from the
background as well as from the gravitational instability.   

\begin{figure}
  \centering
  \includegraphics[width=20pc]{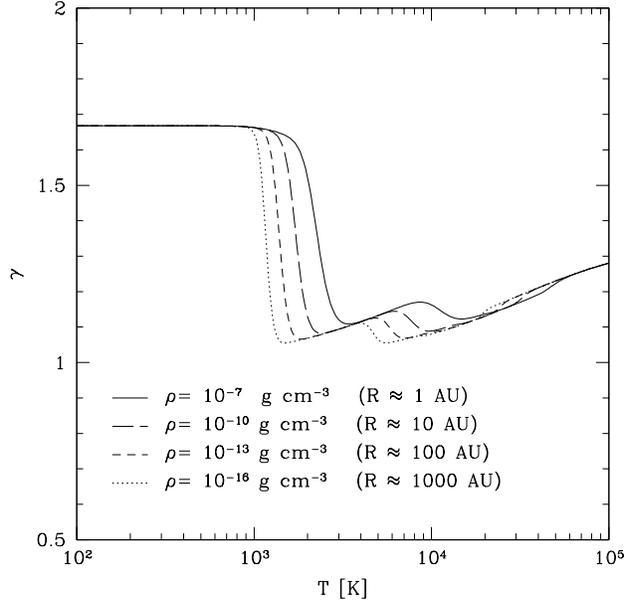}
  \caption{Plot showing the variation of the ratio of specific heats
    $\gamma$ as a function of temperature and density.  The density
    corresponds to the quoted radii for a $Q = 1$ disc about a 1 $M_{\odot}$
    star.  Note that the inverse function for temperature in terms of
    $\gamma$, $T(\gamma,\rho)$ is multi-valued.}
  \label{gamma}
\end{figure}

Since there is a strong dependence on temperature in certain
opacity regimes (see \tref{opacities}) it is important that the equation of
state adequately captures the correct behaviour of both the ratio of specific
heats $\gamma$ and the mean molecular weight $\mu$, as variation in these can
have significant effects on the system overall.  To implement the equation of
state we therefore make the assumption that the gas phase of the disc contains
only hydrogen and helium, in the ratio $70 \colon 30$.  We can make this
assumption because although the metallicity of the disc is important for the
opacity (and thus the cooling), it makes very little contribution to the
equation of state.  Furthermore, the ratio of ortho- to para-hydrogen is
assumed to be held constant at $3 \colon 1$. Following on from the analysis of
\citet{BlackBodenheimer75}, \citet{Stamatellosetal07} produced tabulated
values of $\rho, T, \gamma$ and $\mu$ for this equation of state and it is
these values that we have used throughout.  The variation of $\gamma$ with
both temperature and density is shown in \fref{gamma} -- for the variation of
the mean molecular weight $\mu$ the reader is referred to
\citet{Stamatellosetal07} and \citet{Forganetal09}.  

\begin{figure*}
  \centerline{
    \includegraphics[width=0.35\textwidth]{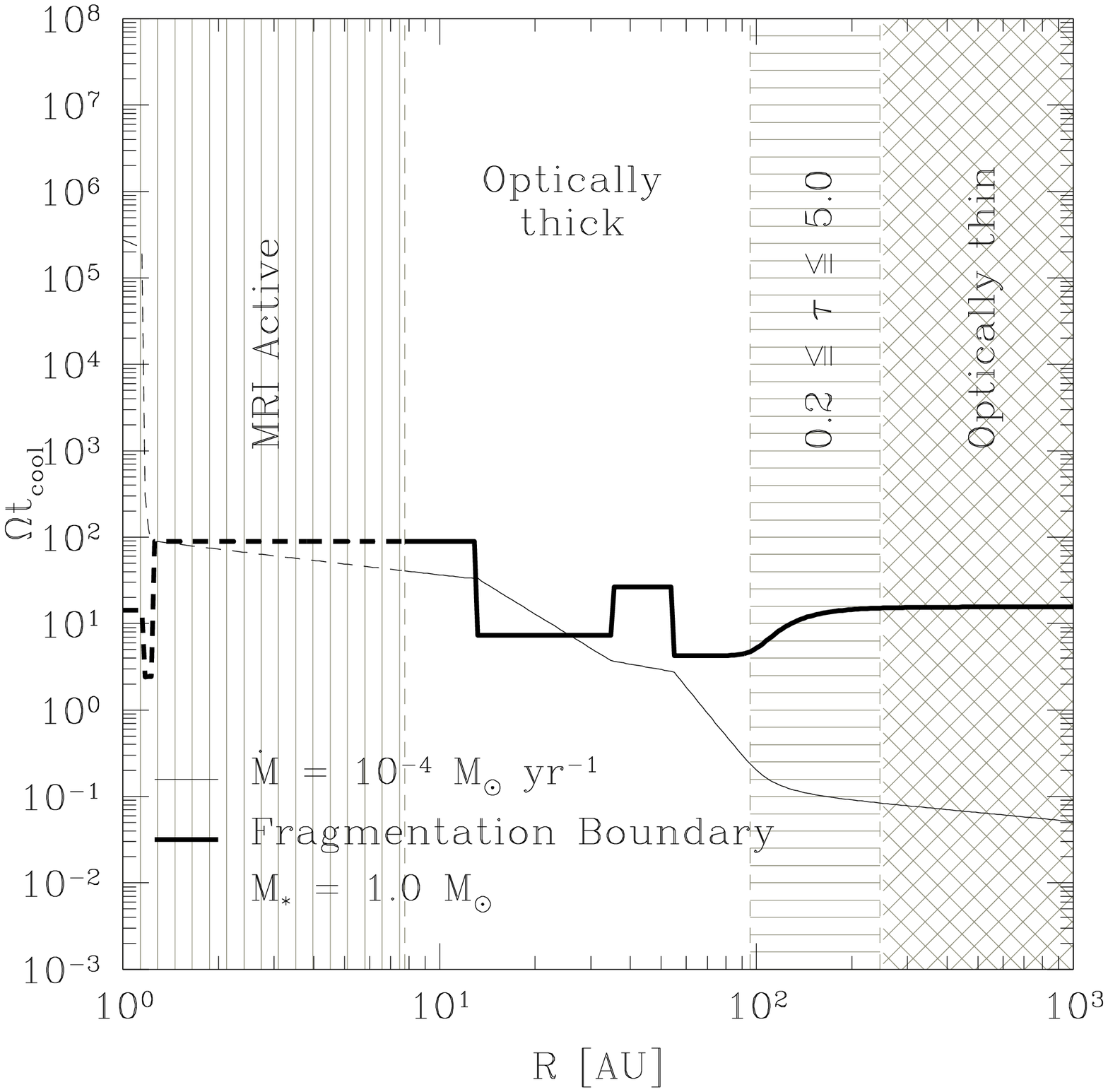}
    \includegraphics[width=0.35\textwidth]{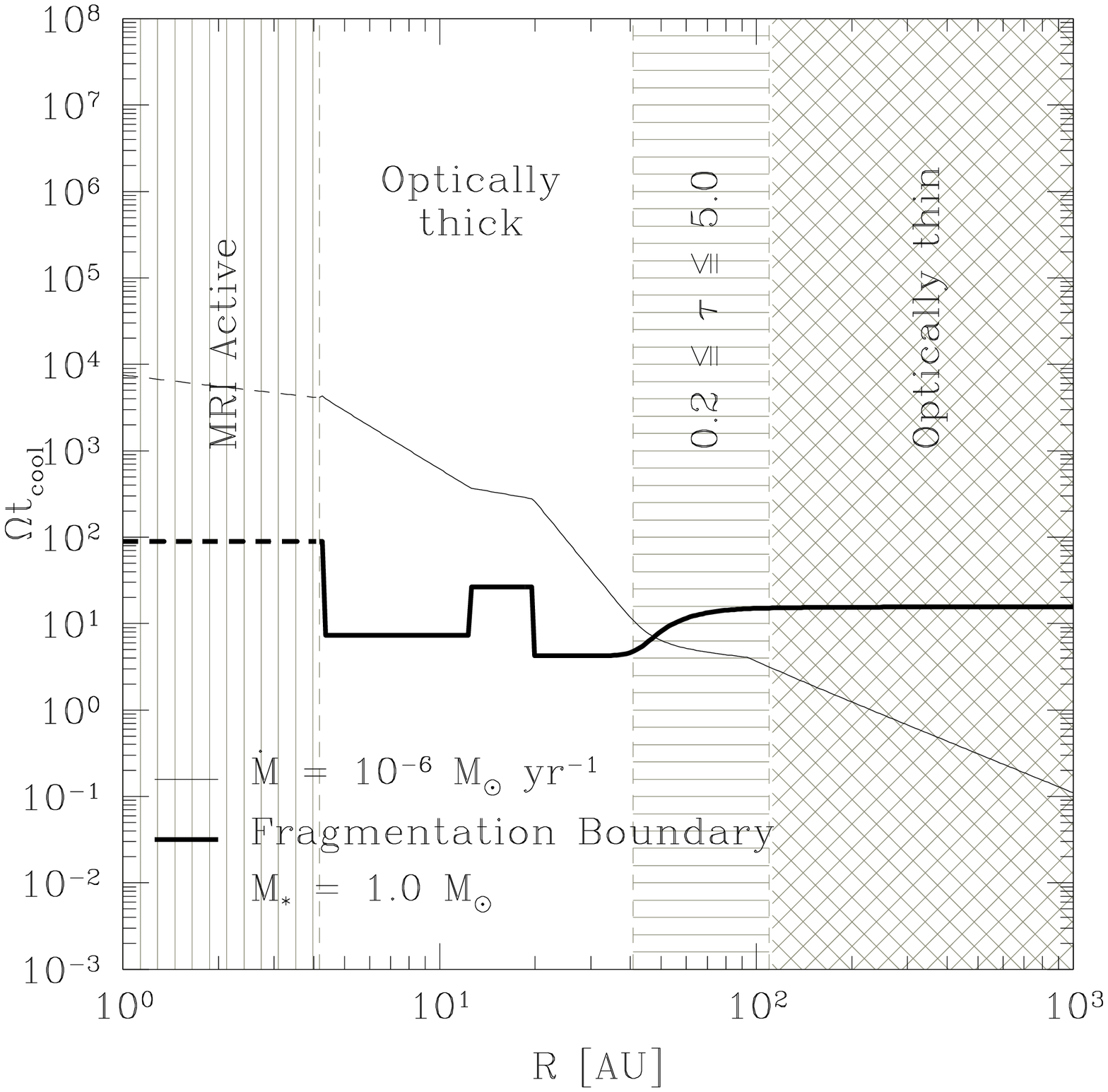}
  }
  \centerline{
    \includegraphics[width=0.35\textwidth]{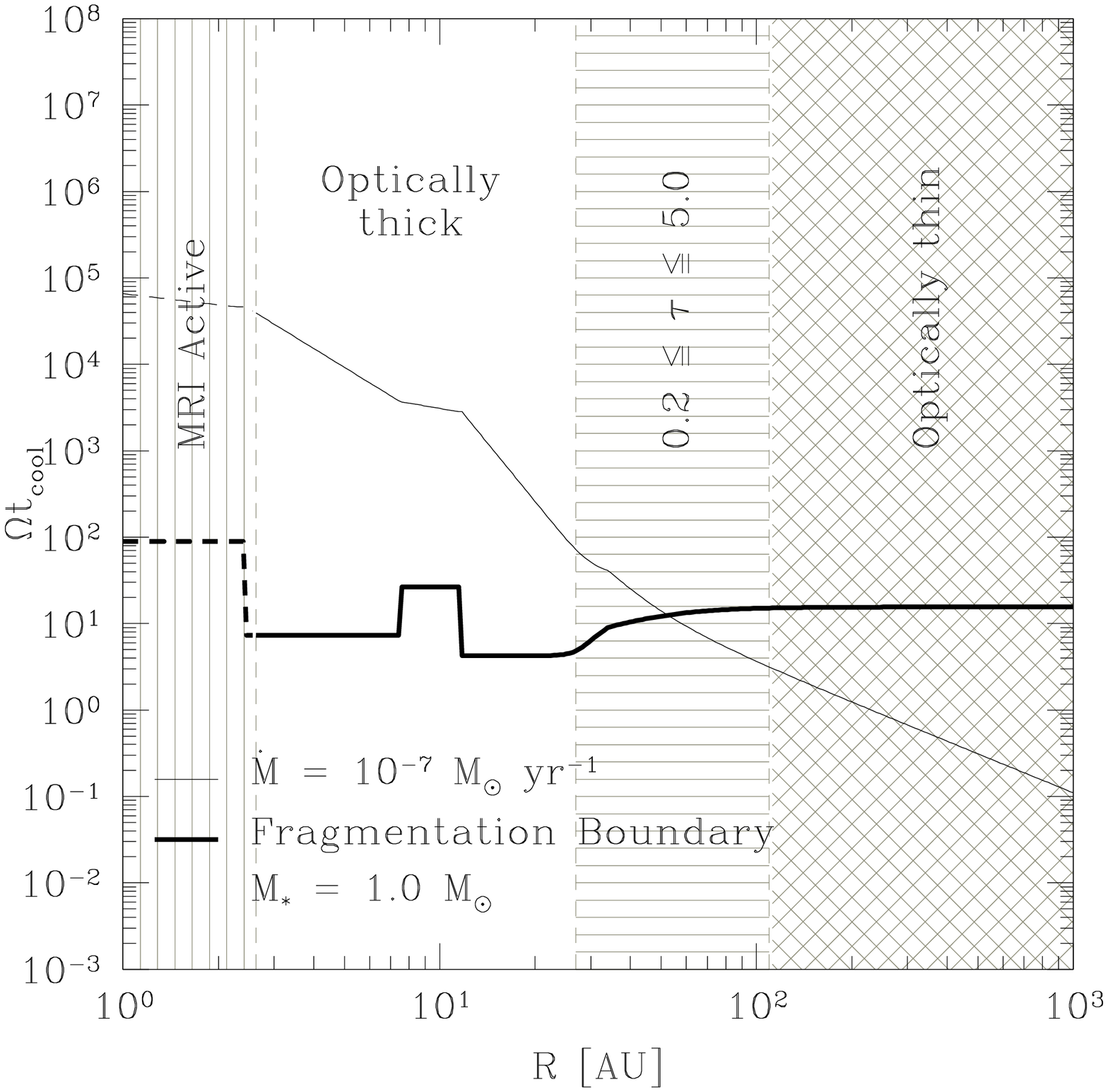}
    \includegraphics[width=0.35\textwidth]{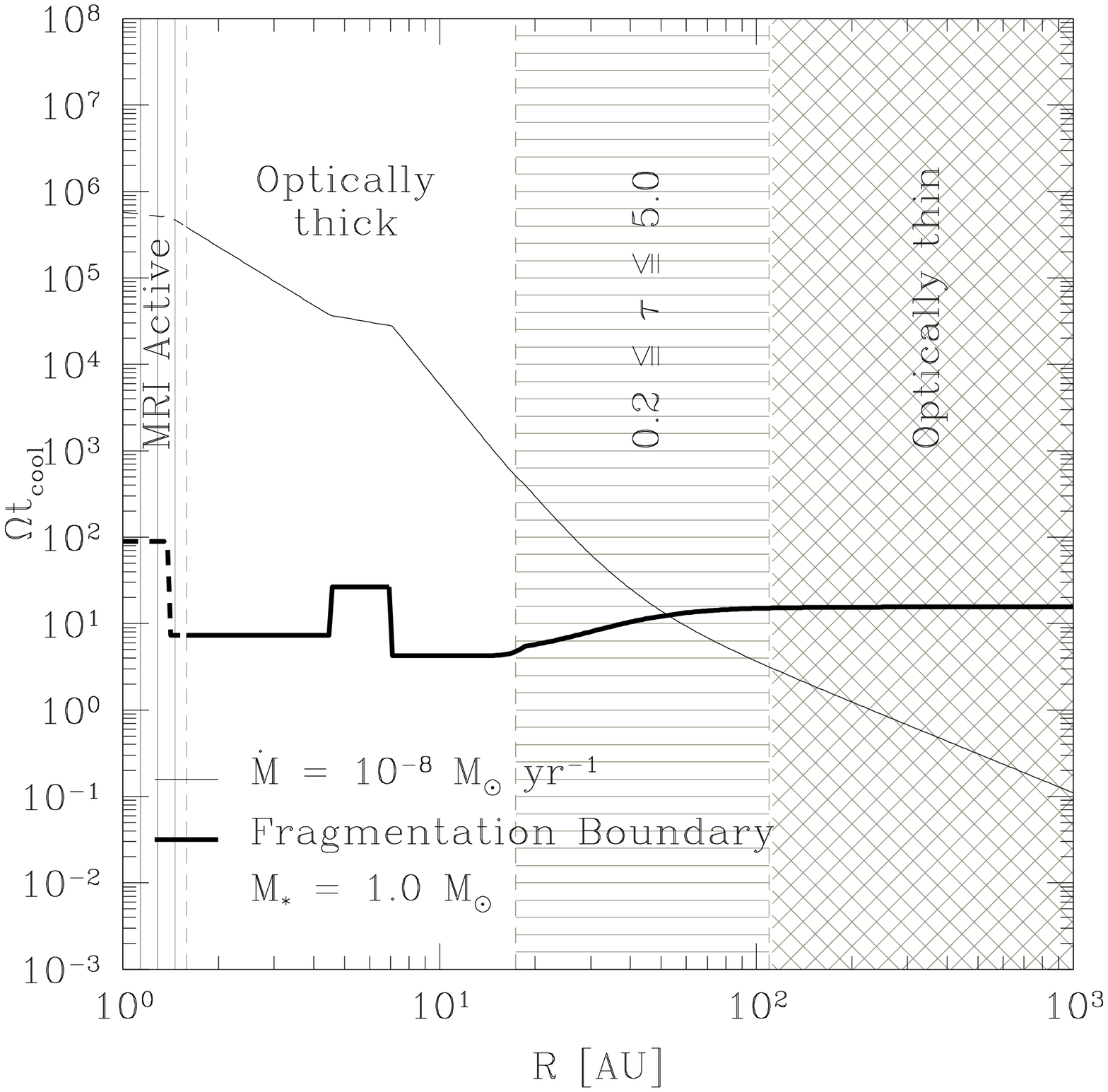}
  }
  \caption{Value of $\Omega \tc$ as a function of radius for accretion rates
    of $10^{-4}$ (top left), $10^{-6}$ (top right), $10^{-7}$ (bottom left)
    and $10^{-8}$ (bottom right) $\msolaryr$, for a disc about a $1
    M_{\odot}$ star.  The unshaded regions are optically thick ($\tau >
    5$), the horizontally shaded areas are transitional ($0.2 < \tau \leq 5$)
    and the cross-hatched regions are optically thin ($\tau < 0.2$).  The
    vertically shaded areas denote regions of the disc that are MRI active.
    The disc is stable against fragmentation wherever the value of $\Omega
    \tc$ is greater than the fragmentation boundary (shown by the heavy solid
    line).  The dotted lines show the values that $\Omega \tc$ and the
    fragmentation boundary would take if the MRI were not active.}   
  \label{AnalyticOmtc}
\end{figure*}

With this tabulated equation of state we can now solve the system of equations
for $\Omega \tc$  for any given values of $Q$, $R$, $\dot{M}$ and $M_{*}$ for
each opacity regime.  For simplicity we assume that the system is marginally
gravitationally stable throughout, such that $Q = 1$.  Furthermore, since we
know the dependence of $\Omega \tc$ on temperature for each of the opacity
regimes, we can use \eref{betan_empirical} (with $\beta_{0}$ = 4.25) to
predict the (average) value of $\Omega \tc$ at which we would expect
fragmentation, the results of which are shown in \tref{opacityfrag}.  Note
that since the value of $\beta_{n}$ depends only on the relative size of the
perturbations in temperature and not on either the mean temperature itself or
the value of $Q$, we do not expect to see any variation in $\beta_{n}$ with
varying $Q$, whereas the value of $\Omega \tc$ will vary with both.  Using
equations \ref{T-MdotR} and \ref{Omtc} we find that the $Q$ dependence of
$\Omega \tc$ is   
\begin{equation}
  \Omega \tc \sim Q^{-1 + 3(a + 1) / (2b - 7)}
  \label{OmtcQ}
\end{equation}
and thus except where $b \approx 3.5$ (such as in the regime where molecular
line cooling dominates the opacity) the effects of $Q$ variation are small.
Nonetheless, in all optically thick cases, the effect of an increase in $Q$ is
to decrease the value of $\Omega \tc$, as can be seen from \tref{opacityfrag}.

\begin{table}
  \caption{Predictions for the fragmentation boundary $\beta_{n}$ for each
    opacity regime in the optically thick case.  The italicised case gives the
    prediction in the optically thin limit for ices, the only regime in
    our models where the disc becomes optically thin.  Note that for large
    positive exponents (such as for hydrogen scattering) the value of
    $\beta_{n}$ becomes undefined.  Note also that where the temperature
    exponent $n$ is positive the regime may become susceptible to thermally
    instabilities.}  
  \begin{center}
    \begin{tabular}{cccc}
      \hline
      \multirow{2}{*}{Opacity Regime} & \multicolumn{2}{c}{Dependence of
        $\Omega \tc$} & \multirow{2}{*}{$\beta_{n}$} \\ 
      & on $T$ & on $Q$ & \\
      \hline
      Ices                    & -- none --     & $Q^{-2}$     &  4.250 \\
      \textit{Ices$^*$}       & \textit{$T^{-5}$} & $Q^{2/3}$ &
      \textit{15.570} \\ 
      Ice Sublimation         &  $T^{-9}$      & $Q^{-8/7}$   & 26.688 \\
      Dust Grains             &  $T^{-3/2}$    & $Q^{-3/2}$   &  7.292 \\
      Dust Sublimation        &  $T^{-26}$     & $Q^{-61/55}$ & 88.296 \\
      Molecules               &  $T^{1}$       & $Q^{-6}$     &  2.427 \\
      Hydrogen scattering     &  $T^{8}$       & $Q^{-9/13}$  & undefined \\
      Bound-Free \& Free-Free &  $T^{-9/2}$    & $Q^{-3/2}$   & 14.297 \\
      Electron scattering     &  $T^{-2}$      & $Q^{-10/7}$  & 8.380 \\
      \hline
    \end{tabular}
  \end{center}
  \label{opacityfrag}
\end{table}

\begin{figure}
  \centering
  \includegraphics[width=20pc]{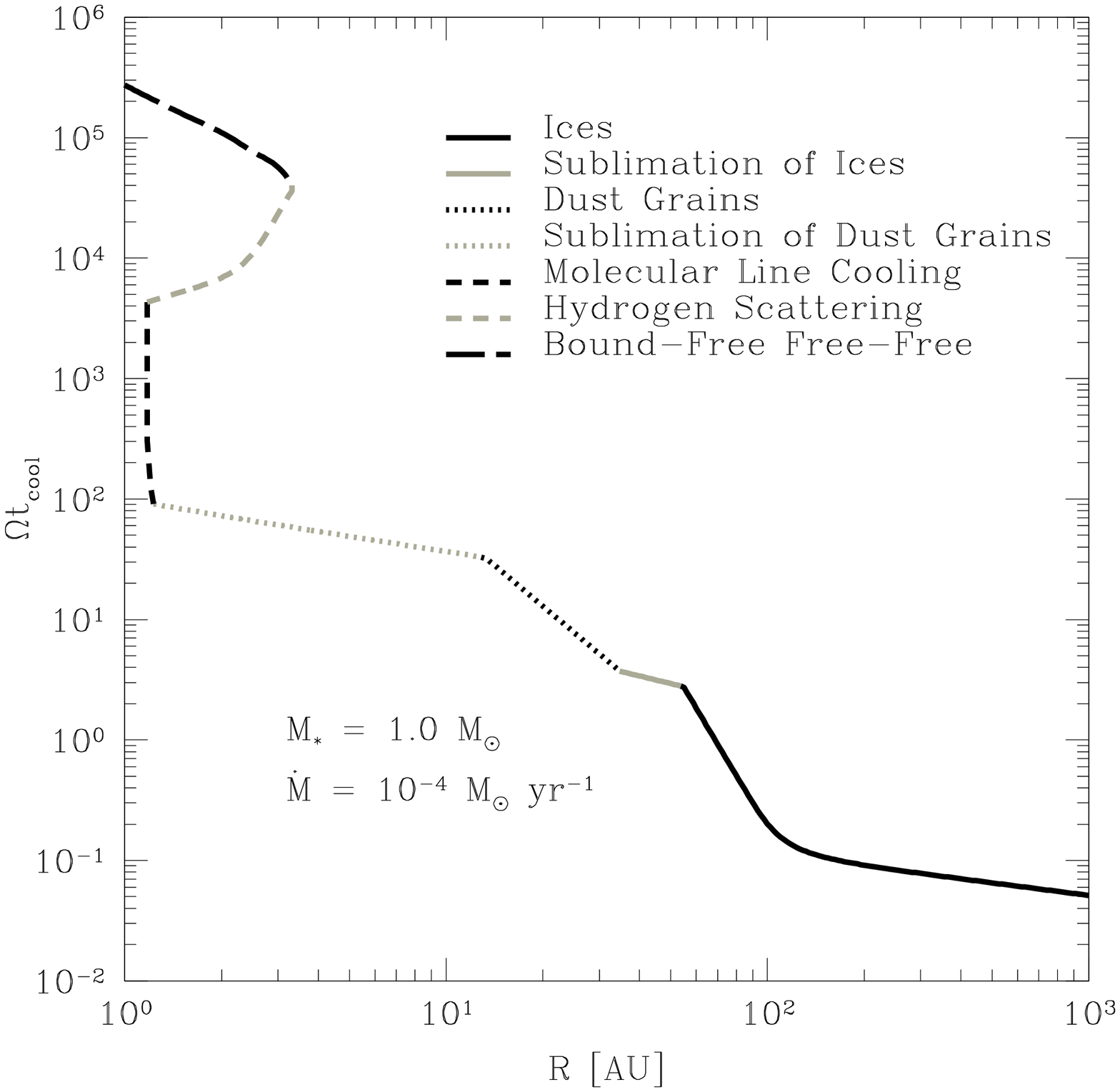}
  \caption{Plot of $\Omega \tc$ for a mass accretion rate of $10^{-4}
    \msolaryr$ indicating the effects of the various opacity
    regimes.} 
  \label{Otcregimes}
\end{figure}

In \fref{AnalyticOmtc} we therefore show the variation in $\Omega \tc$ for a
disc about a $1 M_{\odot}$ protostar as a function of radius at mass accretion
rates of $10^{-4}$, $10^{-6}$, $10^{-7}$ and $10^{-8} \msolaryr$.  (For
completeness, the various opacity regimes are shown in \fref{Otcregimes} for
an accretion rate of $10^{-4} \msolaryr$ -- all other accretion rates are
qualitatively similar.)  From the lower two panels (where the accretion rates
are $10^{-7}$ and $10^{-8} \msolaryr$ for the left and right panels
respectively) we see that at low accretion rates the fragmentation boundary
becomes fixed at approximately $50 AU$, and that this is unaffected by the
transition to the optically thin regime.  This is down to the fact that the
temperature becomes limited below by the background ISM temperature of $10K$,
and is therefore decoupled from the mass accretion rate.  

As the accretion rate rises to $\sim 10^{-4} \msolaryr$ however, the disc
becomes unstable to fragmentation at a wide range of radii due to the increase
in the fragmentation boundary caused by the temperature dependence.  Although
an island of stability exists between approximately $10 - 25 AU$ (where
cooling is dominated by dust grains), all other radii become unstable.  

Note also that at low radii the disc becomes MRI active.  This occurs at radii
from $\sim 1 - 8 AU$ dependent on $\dot{M}$, which corresponds roughly to the
transition to the dust sublimation opacity regime.  For accretion rates of
$\dot{M} \lesssim 10^{-4} \msolaryr$ \fref{AnalyticOmtc} suggests that the disc
will be stable against fragmentation when the MRI is active, as in the absence
of the MRI the value of $\Omega \tc$ would be above the fragmentation boundary.
However, where $\dot{M} \approx 10^{-4} \msolaryr$ the picture is less clear,
as the disc is MRI active whilst simultaneously being unstable to
fragmentation. However, \citet{Fromangetal04} have suggested that where both
instabilities operate the interaction causes the gravitationally-induced
stress to weaken by a factor of two or so, which may stabilise the region
against fragmentation. 

Nonetheless, throughout the range of mass accretion rates investigated here
there are  \textit{no} purely self-gravitating solutions at low radii, as the
MRI is always active.  It is however clear that for radii of $\sim 5 - 50 AU$
the susceptibility to fragmentation of a disc depends strongly on its steady
state accretion rate, and that beyond approximately $50 AU$, with a $10K$
background temperature discs are always unstable to fragmentation.  

\begin{figure}
  \centering
  \includegraphics[width=20pc]{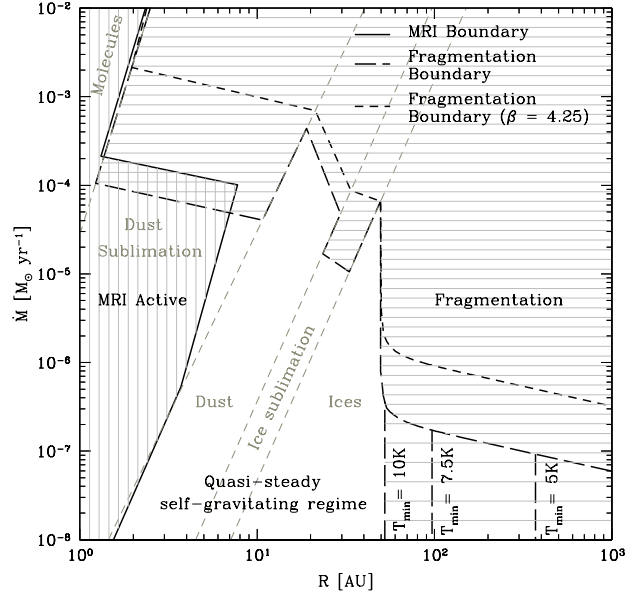}
  \caption{Plot showing the regions expected to be marginally gravitationally
    stable (unshaded), unstable to fragmentation (horizontal shading) and
    unstable to the magneto-rotational instability (vertical shading) in a
    disc about a $1 M_{\odot}$ protostar.  The  cross-hatched regions show
    where where the disc is unstable to both the MRI and fragmentation.  The
    more widely spaced horizontally shaded region to the lower right would
    become unstable to fragmentation if the minimum temperature limit of 10K
    was removed, and the fragmentation boundary moves to the right as the
    minimum temperature is decreased.  The short dashed line corresponds to
    the fragmentation boundary if a fixed value of $\beta = 4.25$ is used
    (cf. \citealt{Clarke09}).}  
  \label{Omtc-MdotR}
\end{figure}

Finally it is useful to see how the fragmentation and MRI boundaries vary as a
function of both $R$ and $\dot{M}$, and this is shown in \fref{Omtc-MdotR}
assuming that as before the central protostar has mass $M_{*} = 1 M_{\odot}$.
Here we have also included the fragmentation boundary in the case where the
effects of temperature perturbations are ignored, i.e. where $\beta = 4.25$ at
fragmentation for all opacity regimes, which allows for comparison with the
work of \citet{Clarke09}.   

\fref{Omtc-MdotR} shows clearly that by including the effects of temperature
perturbations, the mass accretion rate at which fragmentation occurs is
reduced, with an increased effect as the dependence of $\Omega \tc$ on
temperature increases.  As before we note that there is now a region with
$\dot{M} \approx 10^{-4} \msolaryr$ and $R \lesssim 10AU$ where both the MRI
is active and the disc is unstable to fragmentation.  For accretion rates of
$\sim 10^{-5} - 10^{-3} \msolaryr$ there are limited radial ranges where a
marginally gravitationally stable state exists, with regions that are unstable
to fragmentation at both higher and lower radii.

\fref{Omtc-MdotR} also shows how the stability of the disc to fragmentation
varies with the background ISM temperature.  For low mass accretion rates we
see that as the background temperature decreases, the disc actually becomes
stable out to larger radii.  This can be explained as follows:  In the
optically thin case where the cooling is dominated by ices (the regime in
which this phenomenon is found) the value of $\Omega \tc$ is given by 
\begin{eqnarray}
  \Omega \tc & = & \frac{3\mathcal{R}\sqrt{2 \pi G}}{8 \sigma \kappa_{0}}
  \frac{1}{\mu (\gamma - 1)} Q^{1/2} \rho^{1/2} T^{-5} \\
             & = & \frac{3 \mathcal{R} \sqrt{GM_{*}}}{8 \sigma \kappa_{0}}
  \frac{1}{\mu (\gamma - 1)} R^{-3/2} T^{-5},
  \label {Omtc_thin}
\end{eqnarray}
where we have used \eref{rho} to eliminate $\rho$ in \eref{Omtc_thin}.  Hence
at a fixed radius $R = R_{\mathrm{frag}}$, increasing the temperature $T$
decreases $\Omega \tc$ and thereby \textit{destabilises} the disc. Eventually,
for some $T = T_{\mathrm{frag}}$ we reach $\Omega \tc = 15.570$ (from
\tref{opacityfrag}) and the disc becomes unstable to fragmentation.  

From \eref{Omtc_thin} we see that on the fragmentation boundary (where by
construction $\Omega \tc = 15.570$ is constant), $T_{\mathrm{frag}} \sim
R_{\mathrm{frag}}^{-3/10}$.  Now assuming that the temperature at which
fragmentation occurs is at or above the background temperature
(i.e. $T_{\mathrm{frag}} \geq T_{\mathrm{min}}$) then \eref{Mdot} holds, and we
find similarly that the accretion rate at fragmentation
$\dot{M}_{\mathrm{frag}}$ is given by $\dot{M}_{\mathrm{frag}} \sim
T_{\mathrm{frag}}^{3/2}$.  We therefore find that the radius as which
fragmentation occurs increases with decreasing accretion rate such that
$R_{\mathrm{frag}} \sim \dot{M}_{\mathrm{frag}}^{-20/9}$.  Hence,
decreasing the background temperature decreases the accretion rate at which
the disc becomes unstable to fragmentation, and likewise increases the radius
at which this occurs.

Note however that once $T_{\mathrm{frag}}$ is below the background temperature,
(i.e. when $T_{\mathrm{frag}} < T_{\mathrm{min}}$) the disc temperature
becomes decoupled from the accretion rate, and hence all accretion rates below
$\dot{M}_{\mathrm{min}} = \dot{M}_{\mathrm{frag}}(T_{\mathrm{min}})$ are
unstable to fragmentation for radii $R \ge R_{\mathrm{frag}}$.

\section{Discussion and Conclusions}
\label{discussion}

In summary, we have found from controlled numerical experiments with an
imposed temperature dependent cooling law that the effect of temperature
dependence is to increase the value of $\Omega \tc$ at which the disc will
fragment into bound objects.  Furthermore, this tendency to fragment is
greater the more strongly the cooling function depends on the local disc
temperature.  In this respect, this confirms the results of \citet{JandG03},
who likewise noted a markedly increased tendency towards fragmentation in
certain opacity regimes.  This result has been attributed to uncertainty in
the value of $Q$ in the self-regulated state \citep{Clarke09}, equivalent to
uncertainty in the equilibrium temperature in our models.  

However, our results show that this is only one of two mechanisms that affect
the fragmentation boundary, and one that we have been able to account for
\textit{a posteriori} by using effective values of $\beta$ rather than those
input to the simulations.  The other effect is due to the strength of the
intrinsic temperature perturbations about the mean.  In the case where the
cooling law is dependent on temperature, perturbations about the equilibrium
temperature will mean that some fraction  of the gas has a \textit{lower}
value of $\Omega \tc$ than average.  Once this fraction reaches a critical
value, the disc will become unstable to fragmentation.  As the dependence of
the cooling on these temperature perturbations increases, at a given average
value of $\Omega \tc$ the percentage of gas that lies below the critical value
also increases, and thus the average must increase to avoid fragmentation.  

We therefore find that the effect of allowing the cooling function to depend
on the local temperature is to make the disc more unstable to fragmentation, and
we have been able to quantify this variation (see \eref{betan_empirical}).
Combining this with predictions of the temperature dependence of
protoplanetary discs using opacity-based cooling functions, we find that the
fragmentation boundary can be increased by approximately an order of
magnitude in terms of $\Omega \tc$, in close agreement with \citet{JandG03}.
We have also found that the RMS strength of the temperature perturbations can
be correlated to the average cooling strength (see \eref{kempirical}), in a
very similar manner to that found for the surface density fluctuations
\citep{Cossinsetal09}.   

Using these predicted values in analytic models of marginally-gravitationally
stable $Q = 1$ discs with a representative equation of state, we have found
that the susceptibility of such discs to fragmentation into bound objects is
also sensitive to the steady state mass accretion rate, as shown in
\fref{Omtc-MdotR}.  Others have noted that in the optically thick limit where
the opacity is dominated by ices, $\Omega \tc$ is \textit{independent} of
temperature, and thus the cooling rate is determined only by the local
density, itself a function of radius \citep{MatznerL05, Rafikov05, Clarke09}. 
It has therefore been suggested that once the cooling becomes
dominated by ices fragmentation beyond some radius on the order of 100
$AU$ becomes inevitable, and indeed we find that with a background ISM
temperature of $10K$, fragmentation occurs at $\sim 50 AU$ for all accretion
rates below $\sim 10^{-5} \msolaryr$.  

However, if this minimum temperature condition is relaxed, we find that the
change in cooling due to entering the optically thin regime has the effect of
stabilising the disc out to large radii.  (The fact that allowing it to
become cooler actually \textit{stabilises} the disc is due to the fact that in
this regime $\Omega \tc$ increases with decreasing temperature, and thus a hot
disc has a shorter cooling time than a cold one.)  For Class II / Classic T
Tauri objects embedded in a cold medium with accretion rates below a few times
$10^{-7} \msolaryr$, it is therefore possible that extended discs well beyond
100 $AU$ may be stable against fragmentation (they may well be stable against
gravitational instabilitites altogether), and indeed discs with radii of at
least 200 $AU$ have been observed (see for example
\citealt{Eisneretal08}). Nonetheless, discs with accretion rates at the higher
end of the scale ($\dot{M} \approx 10^{-6} \msolaryr$, \citealt{Hartmannbook})
will still be unstable to fragmentation at radii beyond $\sim 50 AU$.  It
should be borne in mind however that in the outer regions of discs where the
surface density is low, non-thermal ionisation (from cosmic rays, X-rays etc)
can trigger the MRI, and this may provide an alternative mechanism for
preventing fragmentation, as shown in \citet{Clarke09}.  

\fref{Omtc-MdotR} also shows another important result, that for accretion
rates between $10^{-8} - 10^{-2} \msolaryr$ discs cannot exist in a
non-fragmenting purely self-gravitating state at radii $\lesssim 2 - 5AU$.  In
this regime discs are either MRI active ($\dot{M} \lesssim 10^{-4} \msolaryr$)
or unstable to fragmentation ($\dot{M} \gtrsim 10^{-4} \msolaryr$).  We also
find that in a narrow band of accretion rates $\sim 10^{-4} \msolaryr$ it is
possible for discs to be both MRI active and unstable to fragmentation,
although the exact interaction of these two instabilities is uncertain (see
\citealt{Fromangetal04}).  It is therefore the case that for steady-state
protoplanetary discs the gravitational instability cannot drive accretion
directly onto the protostar -- either the MRI or the thermal instability must
act at low radii, as has been proposed for FU Orionis outbursts
(\citealt{Armitage01, Zhuetal09}). 

Finally, our results agree with the generally accepted view that planet
formation through gravitationally-induced fragmentation is unlikely to occur
at radii less than 50 - 100 $AU$ \citep{MatznerL05, Rafikov05,
  WhitworthS06, Clarke09, Rafikov09}, although this critical radius varies with
both the mass accretion rate and the background ISM temperature.  Within this
radius the core accretion model remains likely to be the dominant mode of
planet formation.  Outside this radius however, the fragmentation of spiral
arms will produce gaseous planets, a result which matches that of
\citet{Boley09} using a grid-based hydrodynamical model with radiative
transfer -- fragmentation was noted at $\sim 100 AU$ about a $1 M_{\odot}$
protostar.  This result is further corroborated by \citet{StamatellosW08}
whose radiative transfer SPH code suggested a massive disc about a $0.7
M_{\odot}$ protostar would rapidly fragment into planetary mass objects or
brown dwarf companions beyond approximately $100 AU$. Although the mass
accretion rate onto the central object is not stated in either case, we find
that these figures are nonetheless in general agreement with our predictions.


\section*{Acknowledgements}
\label{acknowledgements}
PJC would like to thank Duncan Forgan for providing the equation of state
tables (thereby saving many hours of prospective labour) and also Ken Rice for
helpful discussions.  We would also like to thank Dimitris Stamatellos for a
careful reading of the manuscript. 


\bibliography{Cossins}

\end{document}